\documentclass[aps,floatfix,twocolumn]{revtex4}
\usepackage{graphicx}
\usepackage{epsfig}
\usepackage{amsmath}
\usepackage{graphics}
\usepackage{latexsym}
\begin{document}

\title{Equilibrium interfacial free energies and Turnbull coefficient
for $bcc$ crystallizing colloidal charged sphere suspensions}

\author{Thomas Palberg$^1$, Patrick Wette$^2$, and Dieter M. Herlach$^2$}

\affiliation{$^1$Institut f\"ur Physik, Johannes Gutenberg Universit\"at Mainz, 55099 Mainz, Germany \\
$^2$Institut f\"ur Materialphysik im Weltraum, Deutsches Zentrum f\"ur Luft- und Raumfahrt (DLR), 51147 K\"oln, Germany}

\begin{abstract}
We extend previous analysis of data for the melt-nucleus interfacial free energy, $\gamma$, gained from optical experiments on suspensions of charged colloidal spheres, which crystallize with body centred cubic (bcc) crystal structures. Compiling data from five pure species with different polydispersities and one binary mixture, we find the equilibrium melt-crystal interfacial energy to be considerably larger than the hard sphere reference value. Both this quantity and the entropy of freezing decrease with increasing polydispersity. Moreover, we give a first experimental determination of the Turnbull coefficient for a bcc crystallizing material. The observed value $C_{T, bcc} \approx  0.3$ agrees well with theoretical expectations for bcc systems with short to medium ranged interactions.
\end{abstract}

\maketitle

The kinetics of phase transitions in general depend crucially on the interfacial free energy between the forming nucleus of the new phase and the hosting mother phase. In particular for the case of crystallization this is reflected in the emergent micro-structure and the properties of the formed solid. A deepened knowledge of the interfacial structure and the involved thermodynamics are therefore of great fundamental and practical interest. As early as 1950, Turnbull observed a linear relation between the reduced crystal-melt interfacial free energy of metals, $ \sigma = \gamma n^{-2/3}$, and the molar enthalpy of fusion, $\Delta H_f: \sigma = C_T \Delta H_f / N_A$ \cite{1}. (Here, $\gamma$  denotes the interfacial energy per unit area, $n$ is the number density of atoms, $C_T$ denotes Turnbull´s coefficient and $N_A$ is Avogadro´s number.) While Turnbull used equilibrium values in this relation, Jiang has recently noted that it should also hold for the nucleus-melt interfacial energy at the temperature of nucleation \cite{2}. The simple, yet very useful relation was widely confirmed for metals crystallizing into face centred cubic (fcc) structure, both by experiments \cite{3,4,5} and by computer simulations \cite{6,7}.

Turnbull´s finding was theoretically supported noting that the entropy of fusion for close-packed metals is relatively constant and relies on the configurational entropy of the spherical hard cores \cite{8}. In contrast to similar relations for other phase transitions (e.g. the linear dependence of the liquid-vapour interfacial energy of metals on their enthalpy of evaporation \cite{9}), Turnbull´s coefficient, $C_T$, therefore depends on the crystal structure. For the original data, it was reported to be 0.45 for metals and 0.32 for many non-metals and semi-metals \cite{1}. With improved experimental techniques it was slightly refined, e.g. Kelton reports a value of 0.43 in his compilation of literature data on fcc crystallizing metals \cite{3}. Because most elements and alloys crystallize into dense packed structures, reliable experimental data on other structures are rare \cite{10} and systematic experimental studies are difficult to conduct.
The use of experimental model systems (as done in the present paper) and computer simulations provide alternative access to interfacial free energies and enthalpies of fusion. For the hard sphere reference system, simulations give orientationally averaged values of the reduced equilibrium interfacial free energy  $\sigma_{0,HS} \approx (0.56-0,66)k_BT/a^2$ (where $k_BT$ is the thermal energy and $a$ denotes the particle radius) \cite{11,12,13,14}. Interestingly, Auer and Frenkel reported an increase in interfacial energy from crystallization studies on strongly polydisperse HS systems \cite{15}, however, without accounting for the possibility of fractionation \cite{16}.

For fcc metals more complex interaction potentials are employed, and data compilations suggest Turnbull coefficients around 0.55 \cite{6} or 0.44 \cite{7}. Also the case of other crystal structures was addressed for hard spheres \cite{17,18}, pure elements and alloys \cite{19} or systems interacting via $1/r^n$ potentials with $n \geq 6$ \cite{20}. In particular, for body centred cubic (bcc) structures, interfacial free energies, temperature dependencies and Turnbull coefficients were found to be much smaller than in the fcc case. Simulated metals, e.g. show an average value of $C_{T,bcc} \approx 0.29$ \cite{6}. Very recently, Heinonen et al. studied two bcc crystallizing Yukawa systems comparing state of the art molecular dynamics simulations and theoretical approaches \cite{21}. For these long ranged interaction type, they reported very small interfacial energies well below the HS reference value, a low anisotropy and a $C_{T,bcc}$ of about 0.15.

In the present Letter, we will use experimental data on charged colloidal spheres to systematically study this key parameter of crystallization for the melt-bcc phase transition. Colloidal interactions can be experimentally tailored to display hard sphere (HS), charged sphere (CS) or hard core Yukawa repulsions but also electrostatic or entropic attraction \cite{22}. By combining these, many (structural) phase transitions known from atomic matter can be modeled on a mesoscopic scale including melt crystallization, glass transition, condensation from the vapour phase, or spinodal decomposition, \cite{23,24,25,26}. As in atomic systems, phase transition kinetics are strongly dependent on the composition and structure of the interfacial region, which is here accessible at convenient time scales and constant temperature (due to the presence of the suspending liquid) and by optical methods (due to the colloid-typical length scales) \cite{27}. Thus, in colloidal model suspensions the macroscopic concept of interfacial energy may be explored on the microscopic level and directly compared to theoretical approaches and simulations \cite{28,29}.

For instance, the reduced equilibrium melt-crystal interfacial energy of HS colloidal crystals was determined from the shape of grain boundary grooves analyzed in terms of a capillary vector model to yield  $\sigma_{0,HS} = (0.58 0.05) k_BT/4a^2$ \cite{29} in excellent agreement with most values obtained from scattering experiments \cite{30,31,32} and simulations \cite{10,11,12,13,14}. In addition to an fcc phase, CS also show a bcc phase at low density \cite{33,34,35}. Interestingly, this phase behaviour can be brought into excellent agreement with predictions from computer simulations of the melting line using a Yukawa pair potential and Lindeman's melting criterion \cite{36}. To this end, the CS interaction within the solid phase containing many body terms has to be determined from elasticity measurements and mapped onto an effective Yukawa pair potential \cite{37,38,39}. Also CS melt-bcc phase transition kinetics have been extensively studied. Reaction limited growth velocities were measured \cite{40,41}. Recently, also density dependent nucleus-melt interfacial free energies, $\gamma (n)$, were reported from crystallization experiments \cite{44,45,46,47,48} interpreted in the framework of classical nucleation theory \cite{42,43}. First attempts to deduce a Turnbull coefficient from such data were presented in \cite{49}. In the present paper we compile the available data and considerably extend previous analysis. We report extrapolated equilibrium interfacial energies, enthalpies and entropies of fusion and Turnbull coefficients for six different bcc systems. Moreover we ask for their correlation with e.g. CS surface potentials or system polydispersity.

Systems under consideration are compiled in Tab. I with the corresponding references. All these systems were studied under thoroughly deionized conditions using advanced, continuous ion exchange techniques \cite{50}. The polymer particle species are highly charged with an effective elasticity charge in its saturation limit: $Z_{eff, G} =  \Psi_{eff} a/\lambda_B$ with $\lambda_B = 0.72nm$ being the Bjerrum length, and effective surface potentials of  $\Psi_{eff} = (7.6-9.5)k_BT$ \cite{51,52}. The silica species Si77 was first thoroughly deionized, then NaOH was added up to the equivalence point, yielding a comparably low effective charge and $\Psi_{eff} \approx 6.3 k_BT$. Polydispersities range from low to moderate. In the deionized state all samples, including the mixture, form polycrystalline bcc solids for $n \geq  n_F $.

\begin{table*}[ht]
\begin{tabular}{|p{2,3cm}|p{1cm}|p{1,3cm}|p{1,3cm}|c|p{1cm}|c|c|c|c|p{1,4cm}|}\hline
Sample Batch No. &	Refe-rence & $2a_{TEM}$ $/nm$ & PI & $Z_{eff,G}$ & $\Psi_{eff}$ $/k_BT$ & $n_F/\mu m^{-3}$ & $n_M/\mu m^{-3}$ & $\sigma_0/k_BT$ & $C_{T,bcc}$ & $T\Delta S/kJ$ $/mol$ \\ \hline
PNBAPS68 BASF ZK2168/7387 & [45] & 68 &	0,05 (UZ) &	331$\pm$ 3 & 9.5 & 6.0$\pm$0.3&6.1$\pm$0.3&1.51&0.275& 0.55 \\ \hline
PNBAPS70 BASF GK0748&[53]&70&0.043 (UZ)&325$\pm$3&8.6&1.8$\pm$0.2&2.0$\pm$0.2&1.62&0.364	& 0.45 \\ \hline
Si77 & [49] & 77&0,08 (MIE)&260$\pm$5&6.4&$ > 28 \pm 1$&30$\pm$1&1,13&0,241&0.47 \\ \hline
PS90 Bangs Lab 3012&[46] & 90&0.025 (DLS) & 315$\pm$8 & 8.1&4.0$\pm$0.5 & 7.0$\pm$0.5&4.28&0.316 & 1.36 \\ \hline
PS100B Bangs Lab 3067 &	[46] & 100&0.027 (DLS) & 327$\pm$10 & 7.6 & 4.2 $\pm$ 0.5 & 5.5$\pm$0.2 & 2.75 & 0.235&1.17 \\ \hline
PS90/PS100B 1:1 & [46] & --- & --- & 322$\pm$ 10 & 7.8 & 4.0$\pm$0.5 & 7.2$\pm$0.6 & 2.26 &	0.399 & 0.57 \\ \hline
\end{tabular}\medskip
\caption{Suspension data and results: Lab code and/or manufacturer's Batch No.; references; diameter; polydispersity index (standard deviation normalized by mean diameter with experimental method indicated: UZ: Ultracentrifuge measurements performed by the manufacturer, MIE: USAXS form factor measurements, DLS: Dynamic light scattering); effective charge number from elasticity measurements \cite{37};   effective surface potential \cite{52}; freezing (F) and melting (M) number densities;  extrapolated reduced equilibrium melt-crystal interfacial free energies; bcc Turnbull coefficient; entropy of fusion.}
\end{table*}

\begin{figure}
\includegraphics[width=\columnwidth]{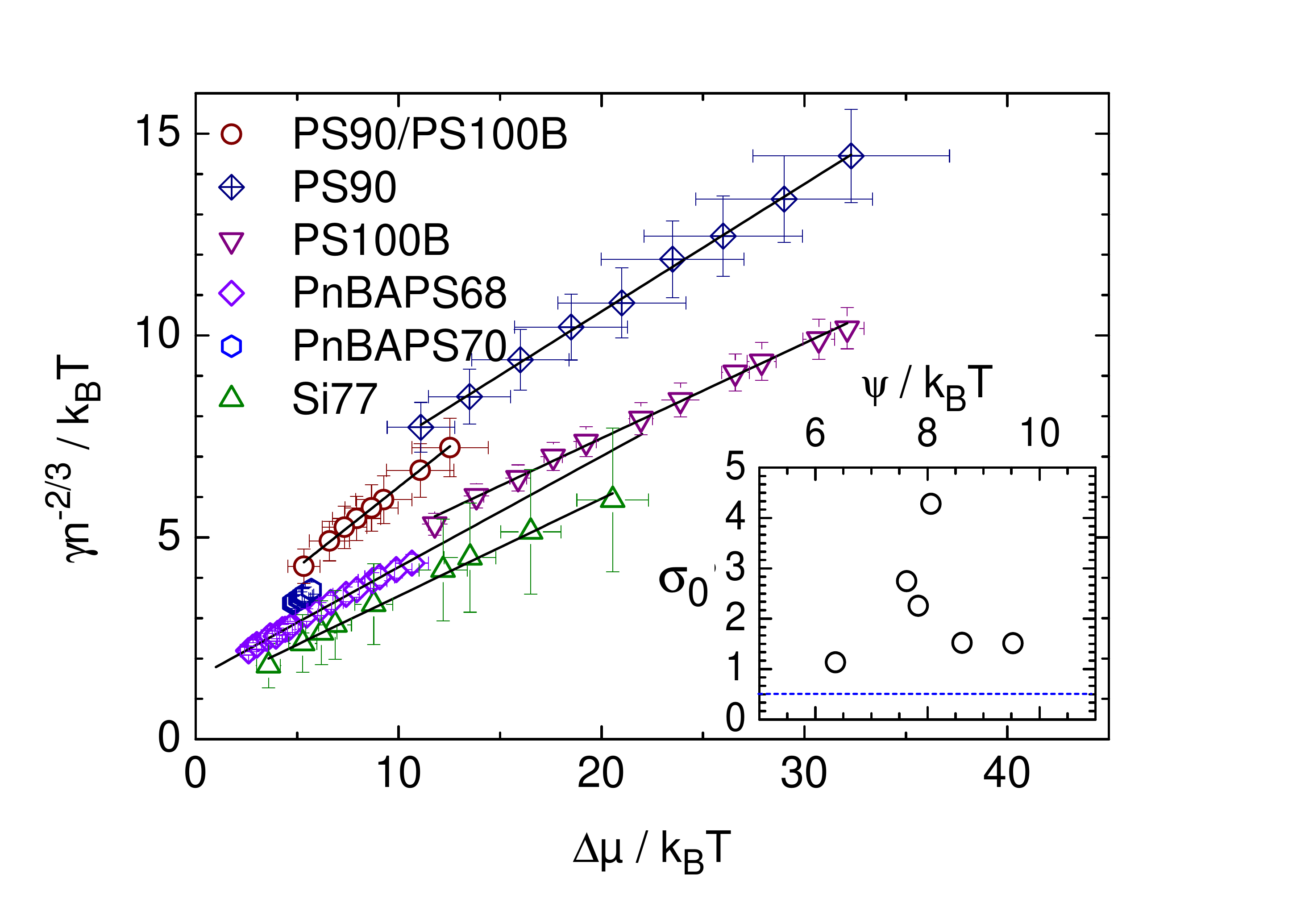}
\caption{\label{fig1} Dependence of the reduced interfacial free energies $\sigma$ on meta-stability  $\Delta \mu$ for the indicated species. Interfacial free energies $\gamma$  are normalized by $n^{-2/3}$ and plotted with uncertainties as quoted from the original literature. Solid lines are fits of $\sigma = \sigma 0 + m \Delta \mu$. Inset: extrapolated equilibrium interfacial free energies $\sigma _0$ in units of the thermal energy for different species versus their effective surface potentials. The dotted line shows the hard sphere reference value.}
\end{figure}

Surface tension data were determined from density dependent crystallization experiments on meta-stable shear melts \cite{44,45,48,52}. To compare the different species, we follow Turnbull \cite{1} and normalize the measured melt-nucleus interfacial free energies by the area taken by a single colloid in the interface, $n^{-2/3} = d_{NN}^2$, where $d_{NN}$ is the nearest neighbour spacing. The resulting reduced quantity is plotted in Fig.~1 versus $\Delta \mu$, as obtained from growth measurements or estimated from the reduced density difference $(n-n_F)/n_F$ using the method of Aastuen et al. \cite{39} with a proportionality constant of 4$k_BT$. The latter results in an enhanced uncertainty in $\Delta \mu$ for PS90 and the mixture. For Si77, graphical evaluation lead to an enlarged uncertainty in $\sigma$. For samples, where $\gamma$ was determined from fits of classical nucleation theory expressions \cite{46} and $\Delta \mu$ was determined from growth experiments following \cite{40}, the statistical uncertainties remain on the order of a few percent.
For all pure species and the mixture, $\sigma$ shows a linear increase of similar slope, $m = (0.25-0.4)$, as  $\Delta \mu$ is increased. Moreover, extrapolation to zero $\Delta \mu$ yields the reduced equilibrium interfacial free energies $\sigma_ 0$ shown in the inset. Their values range between 1.25$k_BT$ and 4.4$k_BT$. Such values are notably larger than the fcc HS reference value and also than the recently observed bcc Yukawa values. Rather, they lie in the range of values for metals.
In the inset of Fig.~1 we plot the extrapolated values versus the corresponding effective surface potentials,  $\Psi$. From the absence of a clear correlation, $\sigma_0$ seems not directly related to the strength of the particle interaction. Graphs of similar scatter are obtained when $\sigma$ is plotted versus the repulsive energy at $d_{NN}$ or the particle size.

In Fig.~2 we plot $\sigma_ 0$ of the pure species versus the corresponding polydispersity. Here, we observe a clearly decreasing trend for increasing polydispersity. Moreover, the value of $\sigma_ 0$ drops below those of the low polydispersity pure components in the case of the binary 1:1 mixture (red bar). This finding seems to be at odds with the observations of Auer and Frenkel \cite{14}. They investigated a hard sphere system and found a strong, nonlinear increase of $\sigma$ with increasing $\Delta \mu$. Absolute values were indistinguishable from the monodisperse reference for polydispersities of 5\%, but significantly larger for larger polydispersities. The range of polydispersities showing an increased $\sigma$ coincided with the range in which later work located segregation effects \cite{15}. Auer and Frenkel's observation may therefore be connected to the neglect of the effects of polydispersity on the phase behaviour. Our systems all show polydispersities of 8\% and lower. Mapping these onto effective hard sphere polydispersities, the values are lowered further and the effect reported by \cite{14} should not become observable. Even our CS mixture does not show any tendency of segregation, rather it crystallizes as substitutional alloy with a spindle type phase diagram \cite{34}.

\begin{figure}
\includegraphics[width=\columnwidth]{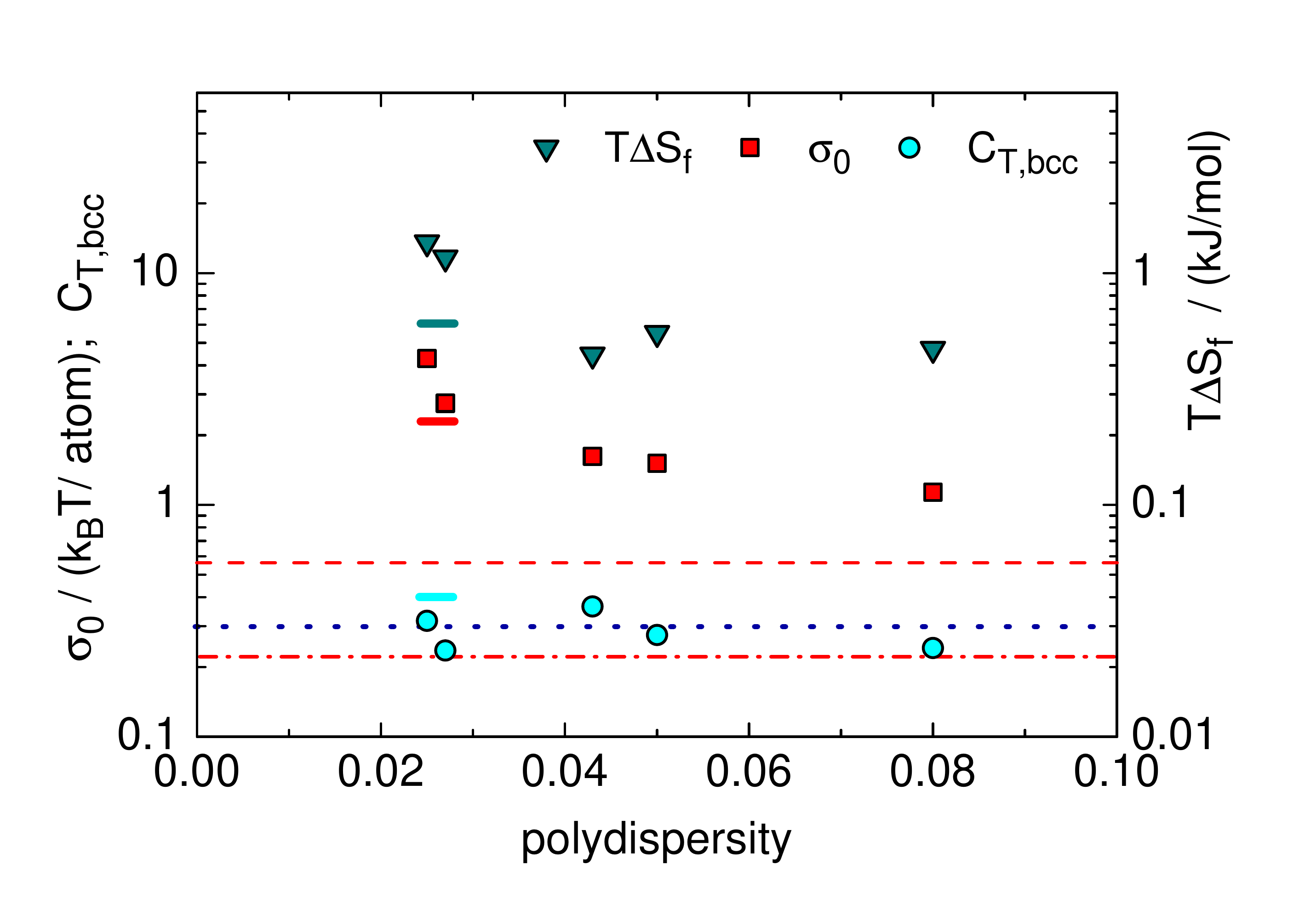}
\caption{\label{fig2} Semilog plot of the dependence of extracted key parameters on system polydispersity. Shown are the bcc Turnbull coefficients (circles) in comparison to the average value observed in simulations (dotted line), the extrapolated reduced equilibrium crystal-fluid interfacial free energies (squares) in comparison to the HS (dashed line) and Yukawa sphere (dash-dotted line) reference values and the entropy of freezing at 298K (down triangles). Symbols denote the five pure components, the mixture is denoted by horizontal bars. While $C_{T,bcc}$ scatters about a value of 0.3 and is not correlated to the particle polydispersity, both $\sigma_ 0$ and $T\Delta  S_f$ show a clear decrease with increasing polydispersity.}
\end{figure}

To obtain the bcc Turnbull coefficient, we next calculate the equilibrium heat of fusion, $\Delta H_f$, per atom. At equilibrium $\Delta G = 0 = \Delta H_f -T\Delta S_f$, and hence $\Delta H_f = T\Delta  S_f$. We apply Turnbull´s relation and calculate  $\Delta H_f / N_A = \Delta \mu  + \sigma_0 /m = \Delta \mu + T \Delta S_f /N_A$, corresponding to a right-shift of the original $\Delta \mu$-dependent data by $T\Delta  S_f /N_A$. The values of (molar) $T\Delta S_f$ are compiled in Tab.~I and also plotted in Fig. 2. Absolute values range between 450 and 1350 J/mol (using T = 298K) which should be compared to the value of bcc metals, which typically is about 10 J mol$^{-1}$ K$^{-1}$ at the melting temperature. As $\sigma_ 0$,  also the entropy of freezing shows a clear decrease with increasing polydispersity. If we follow the arguments given by Laird \cite{8}, we can understand both our observations in terms of a structural change (a deviation from an entropically most favourable configuration) which is more pronounced for the polydisperse crystal than for the polydisperse melt. Consequently also the entropy difference between these and the excess entropy stored in the interface (reflected in $\sigma$) decrease.

We now can construct the Turnbull plot of Fig.~3 plotting the extrapolated equilibrium interfacial free energies (in eV per atom) against the equilibrium enthalpy of fusion (in eV per atom). Values of $C_{T,bcc} =  \sigma _0 / \Delta H_{f,0}$ for each sample are compiled in Tab. I. Within uncertainties we do not find a systematic dependence on polydispersity (c.f. Fig.~2) nor on interaction strength (c.f. Inset Fig.~1). We find an averaged Turnbull coefficient for the data sets of Fig. 4 of $C_{T,bcc} = 0.31 \pm 0.03$. Using only those low-uncertainty data with  µ derived from growth measurements, we obtain a slightly lower value of $C_{T,bcc} = 0.25 \pm 0. 02$. For comparison, we also show literature data for metal systems obtained from experiment (squares \cite{3}) and simulations (diamonds \cite{6}) for fcc metals. Our data points for bcc colloids clearly fall below the latter. Hence, our experimentally determined bcc Turnbull coefficients are considerably smaller than previously reported Turnbull coefficients for fcc systems: $C_{T,fcc,exp} = 0.43$ (dotted line) and $C_{T,fcc,sim} = 0.55$ (dashed line). In fact, our values are much closer to those from simulations of bcc metals (open triangles \cite{6}) which yielded $C_{T,bcc,sim} = 0.29$ (dash-dotted line) \cite{6}. We also compare to the Turnbull coefficients corresponding to the work of Heinonen et al. (down triangles \cite{20}) using the values of Meier and Frenkel \cite{55} for $\Delta H_f$. The corresponding $C_{T,bcc,\textrm{Yukawa}}=0.15$ is much smaller than our experimental value and any other reported $C_{T,bcc}$ including those obtained for simulations on bcc metals using embedded atom potentials. Further, also the $C_{T,fcc}$ values obtained by these authors for Yukawa systems are much smaller than for fcc metal systems and HS, which themselves were found in good agreement with published experimental and simulation data \cite{56}. We think that this systematic shortfall may be connected to the long-rangedness of the Yukawa potential. In the experiments, the repulsive interaction is typically truncated at the nearest neighbour shell due to many body terms in the potential of mean force \cite{57} and the effective pair potential becomes much shorter ranged. Possibly therefore, a pure Yukawa potential is not particularly useful to predict the interfacial free energy and Turnbull coefficient of CS colloids.

\begin{figure}
\includegraphics[width=\columnwidth]{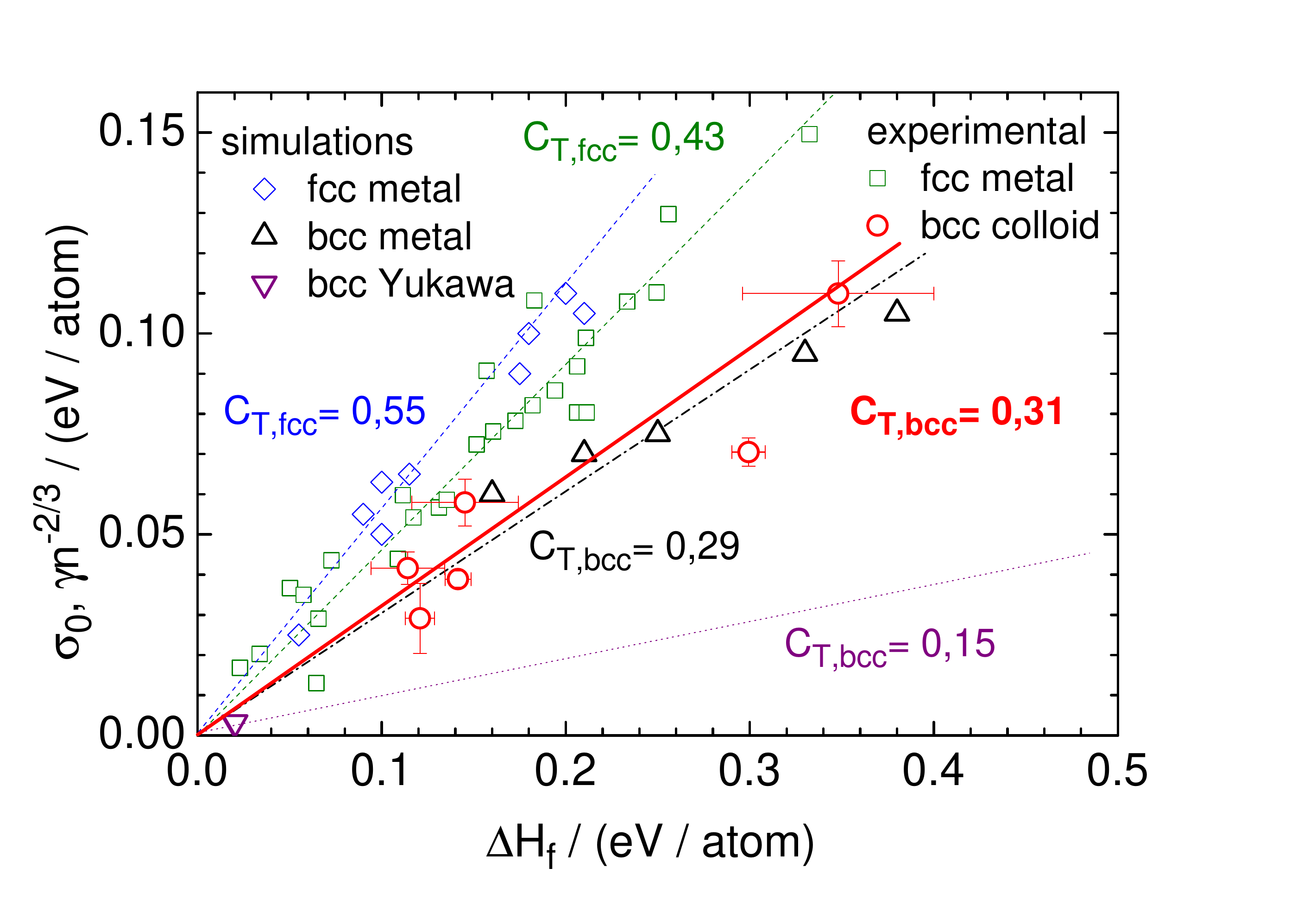}
\caption{\label{Fig3} Turnbull plot of the reduced equilibrium interfacial free energy versus the equilibrium enthalpy of fusion. Shown are our data for bcc crystallizing colloids (circles), simulation data for bcc crystallizing metals (up triangles), bcc crystallizing Yukawa spheres (down triangles) and fcc crystallizing metals (diamonds) \cite{6} as well as experimental data for fcc crystallizing metals (squares) \cite{3}. Lines correspond to the indicated average values of Turnbull coefficients as quoted from \cite{6} and \cite{3} for simulation and experimental data, respectively. For our data we find an average value of $C_{T,bcc} 0.31 \pm 0.03$ (thick solid line).}
\end{figure}

Finally, we show in Fig.~4 that colloids allow to determine non-equilibrium Turnbull coefficients over a much larger range than accessible for  metals. This, of course is due to the colloid-specific possibility to shear melt our samples. In fact, defining an effective transition temperature $T_{eff} = k_BT /V(d_{NN}) $ with $V(d_{NN})$ denoting the pair energy at the nearest neighbour distance, relative ``undercoolings'' of $\omega   = (T_{eff,0}-T_{eff})/T_{eff,0} \approx  0.15$ can easily be reached. Under such conditions the enthalpies of fusion are up to three times larger than for metals at maximum undercooling. Still our experimentally determined reduced interfacial free energies increase nearly linearly across the whole range with possibly a slight deviation towards sub-linear behaviour at large $\Delta H_f$. Clearly, investigations at even larger undercooling are desirable to test such a trend which could e.g.~indicate an increased influence of polydispersity for smaller inter particle distances. Within experimental uncertainty, however, Fig.~4 systematically confirms Jiang's point, that Turnbull's relation should hold for both equilibrated and meta-stable conditions.

\begin{figure}
\includegraphics[width=\columnwidth]{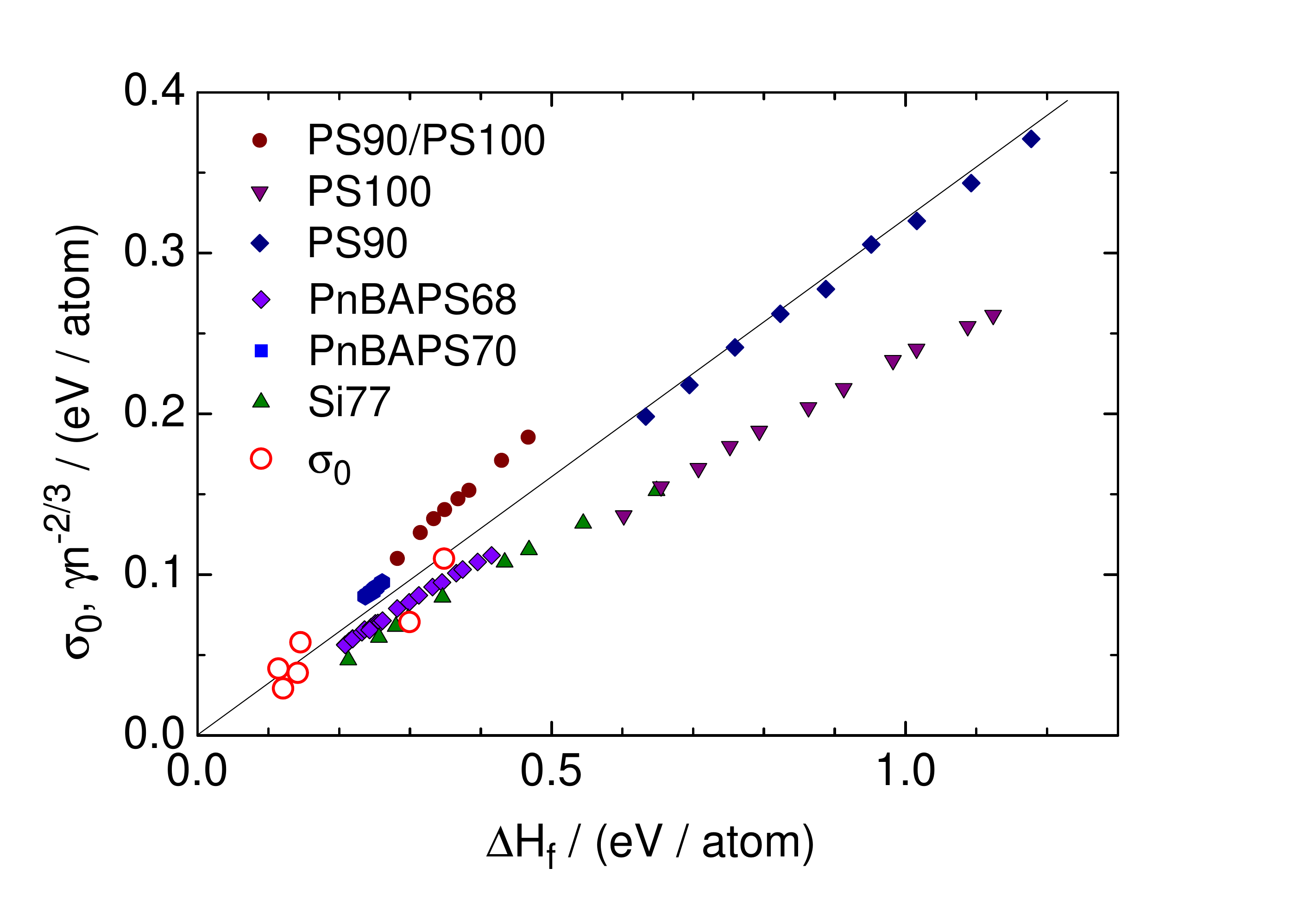}\caption{\label{fig4} density dependent reduced interfacial free energies plotted versus the corresponding enthalpies of fusion for the different colloidal species as indicated. Open symbols denote extrapolated $\sigma_ 0$, the solid line gives the best fit linear dependence. All data arrange neatly on nearly straight lines through the origin.}
\end{figure}
Concluding, we have considerably extended previous data analysis for six bcc crystallizing experimental systems. For the first time we were able to systematically test obtained equilibrium interfacial free energies and Turnbull coefficients for their dependence on solid structure, interaction strength and the colloid specific polydispersity. Like in simulation, $C_T$ for our open crystal structures are much smaller than those observed for close packed structures. Further, we did not observe a particular dependence of Turnbull´s relation on the interaction strength. However, comparison to data obtained from simulations on hard or very long ranged potentials shows a clear dependence of $C_T$ on the type of interaction. Our findings for the interfacial free energy applied to both fluid-crystal equilibrium and melt-nucleus values obtained at large meta-stability. Finally the colloid-specific polydispersity was observed to have a pronounced lowering influence on the melt-crystal interfacial free energy at least for the here analyzed CS systems. We hope that our studies encourage further simulations based on potentials accounting for many body effects.

We are pleased to thank J. Horbach, Th. Speck, and H. L\"owen for fruitful discussions and H. J. Sch\"ope for some critical remarks. Financial support by the DFG (Pa459/16,17 and He1601/24) are gratefully acknowledged.

\end{document}